# Structural and environmental monitoring of tracker and vertex systems using Fiber Optic Sensors

David Moya[1] and Ivan Vila[2] *

1 – IFCA– High Energy department
Avnda Los Castros S/N 39005 Santander – Spain

2 – IFCA– High Energy department
Avnda Los Castros S/N 39005 Santander – Spain

Fibre optic sensors (FOS) are an established technique for environmental and deformation monitoring in several areas like civil engineering, aerospace, and energy. Their immunity to electromagnetic and magnetic fields and nuclear environments, its small size, multiplexing capability and the possibility to be embedded make them an attractive technology for the structural and environmental monitoring of collider particle physics experiments. Between all the possible Fibre Optic sensors FBGs (Fiber Bragg Grating) seems to be the best solution for HEP applications. The first step was to characterize FBG sensors for it use in High Energy Physics environment. During last two years we have checked the resistance of the Fibre Bragg Grating sensors to radiation. Two irradiation campaigns with protons have been done at CNA [1](Centro Nacional de Aceleradores).

In the near future these sensors are being planned to be used in detectors (the closest one Belle II. ). Several work on integration issues in Belle II PXD-SVD, and checking for environmental and deformation monitoring in the detectors inner part has been done.

## 1 Introduction to FBG sensors

Structural and health monitoring using FOSs ( Fiber Optic sensors) has had a fast development during last years. Several techniques based on Fiber Optics has been developed. FOS have several advantages with respect to conventional strain sensors:
- *Immunity against high electromagnetic fields, high voltage, and high and low temperatures*
- *Magnetic field immunity*
- *Light weight and small space (miniaturized)*
- *Flexible*
- *Low thermal conductivity (same as the fiber)*
- *Low signal loses and long range signal transmission*

Between all FOS sensors, FBG sensors are the best option for its application at HEP. Their main advantages are:

- *Multiplexing capability ( sensor network)*
- *Can be embedded in composite structures*
- *Signal is wavelength encoded*
- *Reasonable cost*
- *Can be used in high and low temperatures.*

Those are the reasons why we are working in the characterization of the FBG sensors for its use in HEP

## 1.1 FBGS working principle.

An FBG sensor is a grating of reflective index periodic variation in the core of a fibre. They are normally manufactured by "inscribing" or "writing" systematic variation of refractive index into the core of an optical fibre using an UV laser and a mask or by two ultraviolet lasers interferences (figure 1).

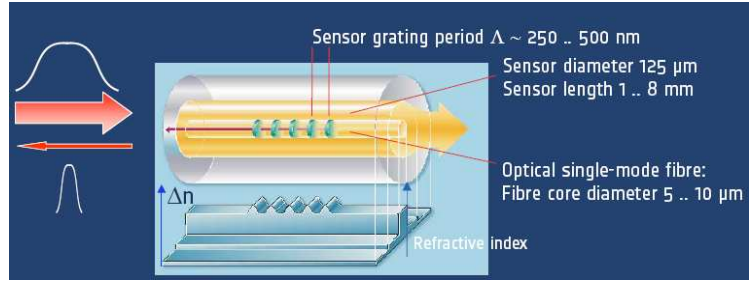

*Figure 1: FBG sensor*

When a broadband light is injected through the fiber, the FBG sensor reflect the light at a defined wavelength (Bragg wavelength). This reflected wavelength is dependent of the grating period and the refractive index change:

$$\lambda = 2n_e \Delta$$

Where $\lambda$ is the reflected wavelength of the FBG sensor, $n_e$ is the refractive index of the grating, which changes with temperature (Thermo-optic coefficient) and with the deformation (photo-elastic coefficient). Finally, $\Delta$ is the period of grating that change with mechanical and thermal deformations.

So, as the sensor reflected Bragg wavelength is dependent of temperature and strain, these FBG sensors can be used to measure these two parameters. The thermal and strain sensibility of an FBG sensor is defined in the next way:

Strain sensibility $\quad \Delta \lambda_b / \lambda = \varepsilon(1-p) \quad$ ( $1\ pm = 1\ \mu\varepsilon$)

Temperature sensibility $\quad \Delta \lambda_b / \lambda = \Delta n[T]/n + \Delta T.\alpha.\varepsilon(1-p) \quad (10\ pm = 1ºC)$

Where $\varepsilon$ is the strain of the fiber, $p$ is fiber photo-elastic coefficient, $\Delta n[T]/n$ is the change of the refractive index of the sensor due to the temperature (thermo-optic coefficient), $\Delta T$ is the temperature change and $\alpha$ is the thermal expansion coefficient.

# 2 Radiation Qualification of FBGS

During last twenty year's several works has been done to study the FBG sensors behavior under nuclear environment, most of its for Gamma radiation [2]. For assure the feasibility for use FBG sensors in HEP their sensibility to proton, gamma and electrons radiation must be studied, experimentally and in the

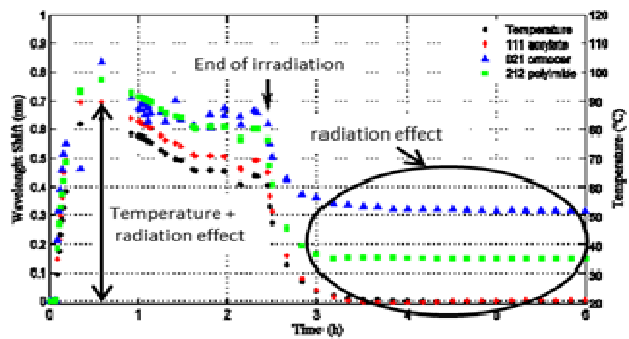

*Figure 2: FBG wavelength change during and after irradiation*

*bibliography. The aim is to find a radiation hard sensor for its application in HEP.*

*A first active irradiation campaign was done in October 2010 at CNA using a 13.5 MeV proton beam up to a fluence of $3.3 \times 10^{15}$ protons/cm$^2$ (total absorbed dose of 15 MGy) . The aim was to find the main parameters affecting the signal of the sensors. During the irradiation, the fibers where in contact with the support, heated due to the proton beam, not allowing disentangling the temperature and radiation effect on the fiber response. Once the proton beam was switched off, the aluminum support got cold and the remaining change in the reflected wavelength of the sensors was a direct effect of the irradiation (Figure 2).*

| Sensor code | Coating | Manufacture | Type | Reflectivity (dB) | Peak Shift (pm) |
|---|---|---|---|---|---|
| 111 | acrylate | Conventional | Type I | 0.42 | -3 |
| 112 | acrylate | Conventional | Type I | 0.15 | -14 |
| 211 | polyimide | Drawtower | Type I | 0.44 | 10 |
| 212 | polyimide | Drawtower | Type I | 6.05 | 140 |
| 213 | polyimide | Drawtower | Type I | 6.26 | 173 |
| 214 | polyimide | Drawtower | Type I | 7.16 | 175 |
| 221 | polyimide | Drawtower | Type I | 7.56 | 158 |
| 222 | polyimide | Drawtower | Type I | 6.92 | 175 |
| 223 | polyimide | Drawtower | Type I | 6.92 | 173 |
| 021 | Ormocer | Drawtower | Type I | 6.23 | 308 |
| 022 | Ormocer | Drawtower | Type I | 8.80 | 379 |
| 023 | Ormocer | Drawtower | Type I | 5.00 | 334 |
| 03 | Ormocer | Drawtower | Type II | 0.01 | 206 |
| 04 | Ormocer | Drawtower | Type II | 0.03 | 307 |
| 05 | Ormocer | Drawtower | Type II | 0.21 | 359 |

*Table 1: 2010 FBG irradiation results*

*This first irradiation results (table 1) allowed us to establish that the main effect concerning the wavelength shift was the type of the fibres coating, while the reflectivity change of the sensor is more affected by the manufacture procedure and the type of fibre used for the FBG manufacture. The results of this first irradiation where presented at RADECS 2011 [3]*

## 2.1    Second Irradiation campaign September 2011 motivation

*A second irradiation campaign was done to be able to disentangle the wavelength shift due to Ionization effects and radiation induced temperature change effect. In order to archive this, a new support was used (figure 3).*

*The other main reason was to irradiate the fibers up to a dose of 10 Mrad, the one expected in the tracker of Belle II experiment.*

## 2.2 Second Irradiation campaign Set-up and development

*In this second irradiation nine fibres where irradiated with three different coatings (ormocer, acrylate and polyimide). They were written ussing different techniques. Some of the sensors were uncoated or embedded in Carbon fibre / glass fibre composite layout. In table 3 are listed the fibres and sensors used for this second irradiation. Each fibre have three sensors, two of them inside irradiation area and the third one outside to measure temperature change in the irradiation bunker (as reference)*

*The fibres were mounted vertically (figure 3) and free in the air, locked in the top of the structure with adhesive tapes and guided by one millimeter grooves in the support, to ensure the position of the fibres during irradiation without constraining them.*

*The irradiation was done in steps, in order to measure the evolution of the wavelength and reflectivity change with the absorbed dose. Five steps where defined at a total absorbed dose of 0.5, 1, 2, 5 and 10 Mrads.*

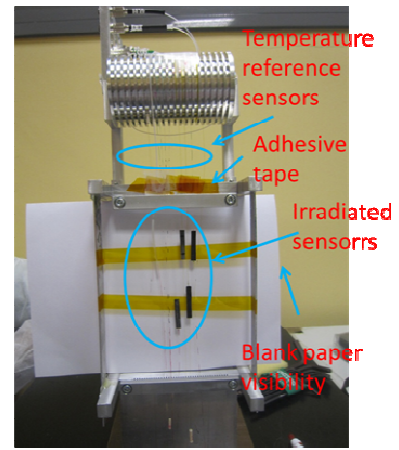

*Figure 3: 2011 irradiation Set-Up*

| Sensor | Coating | Manufacture | Type | Status | Peak shift (pm) | Attenuation (dB) |
|---|---|---|---|---|---|---|
| 111 | Acrylate | Draw Tower | Normal | Referencia | Not applicable | -0,0221 |
| 112 | Acrylate | Draw Tower | Normal | irradiated | 30,05 | -1,4867 |
| 113 | Acrylate | Draw Tower | Normal | irradiated | 28,33 | -1,5178 |
| 121 | Acrylate | Draw Tower | Normal | Referencia | Not applicable | -0,0308 |
| 122 | Acrylate | Draw Tower | Normal | irradiated | 28,91 | -1,33 |
| 123 | Acrylate | Draw Tower | Normal | irradiated | 28,44 | -1,3697 |
| 211 | Ormocer | Draw Tower | Normal | Referencia | Not applicable | -0,1476 |
| 212 | Ormocer | Draw Tower | bare | irradiated | 56,98 | -0,216 |
| 213 | Ormocer | Draw Tower | Normal | irradiated | 49,33 | -1,4908 |
| 221 | Ormocer | Draw Tower | Normal | Referencia | Not applicable | -0,0132 |
| 222 | Ormocer | Draw Tower | Normal | irradiated | 52,69 | -1,2905 |
| 223 | Ormocer | Draw Tower | embedded | irradiated | 48,38 | -1,4163 |
| 231 | Ormocer | Draw Tower | Normal | Referencia | Not applicable | -0,003 |
| 232 | Ormocer | Draw Tower | embedded | irradiated | 39,59 | -1,3326 |
| 233 | Ormocer | Draw Tower | Normal | irradiated | 56,63 | -1,4282 |
| 311 | Polyimide | Normal | Normal | irradiated | 55,68 | 0,0048 |
| 312 | Polyimide | Normal | embedded | irradiated | 74,81 | 0,2936 |
| 313 | Polyimide | Normal | Normal | Referencia | Not applicable | -0,1602 |
| 321 | Polyimide | Normal | bare | irradiated | 49,5 | 0,1036 |
| 323 | Polyimide | Normal | bare | No active measure. | not measured | |
| 331 | Polyimide | Normal | Normal | Referencia | Not applicable | -0,0692 |
| 332 | Polyimide | Normal | Normal | irradiated | 52,38 | -0,0185 |
| 333 | Polyimide | Normal | embedded | irradiated | 60,73 | -0,1669 |
| 411 | Acrylate | Normal | Normal | irradiated | 16,04 | 0,0353 |
| 412 | Acrylate | Normal | Normal | irradiated | 13,34 | 0,1692 |

*Table 3: 2011 Irradiated sensors and results*

## 2.3 Second Irradiation campaign: Preliminary Results.

*Figure 4 shows the change of the sensor wavelength with time during irradiation of tow polyimide coated sensors, one of them embedded. In the Figure can be seen 5 peaks, corresponding with the five steps of irradiation. There are three areas where the signal of the sensor becomes noisy (before first step, between first and second step, and after last step) due to the opening of the door of the irradiation bunker, which introduced temperature changes.*

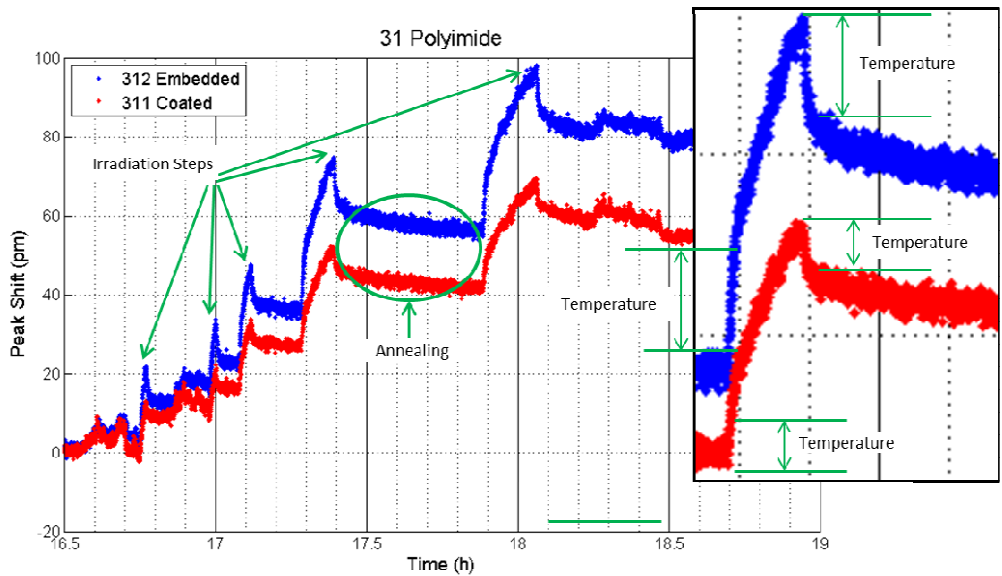

Figure 4: Sensors Peak Shift with time during irradiation

*From this figure we can conclude that the maximum temperature reached by the fibres was of 3 ºC. At the beginning of each irradiation step is observed a lineal increment of the wavelength caused the fibre temperature change and radiation effect. Then the wavelength increment become slower with time (the fibre has reached the thermal stability) and the only effect affecting the sensor would be the radiation.*

*After each irradiation step a similar effect is observed. There is a big and fast change of the wavelength which corresponds to a thermal stabilization of the sensor, and after this fast change there is a slower effect of annealing of the sensor. This annealing of the sensor must be studied in more detail.*

*The other interesting result to be checked is the change of the wavelength*

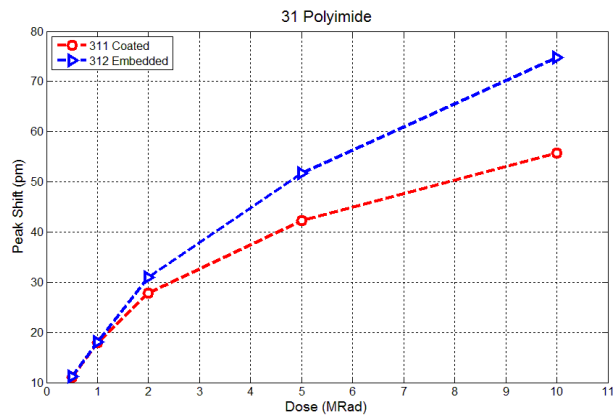

Figure 5: Sensors Peak Shift change with absorbed dose

*with respect to the absorbed dose. In the picture 5 it can be seen the wavelength change of the sensor with respect to the absorbed dose for two polyimide sensors. The sensibility of the sensor to radiation is lower at higher absorbed doses, because it starts to saturate. The main conclusion is that the FBG sensors could burn up to the saturation absorbed dose to reduce or eliminate the effect of the radiation in the FBG response. In table 3 can be seen the final effect of irradiation in the sensors.*

## 3   Environmental measurements on the thermal characterization set-up of Belle II vertex detector

*Taking into account that the FBG sensors will be used for environmental, deformations and displacement monitoring at Belle II vertex detector , it was found useful to do some real measurements in the thermal characterization set-up of Belle II vertex detector at Valencia.*

*The main goal was to measure temperature, deformation and vibrations of several parts of the set up in Valencia´s thermo-mechanical mock-up during cooling cycles. The objective was to identify first order issues on the use of FBG sensors for monitoring.*

### 3.1   Belle II vertex detector Thermo-mechanical set up

Belle II thermo-mechanical set-up is a PXD  Belle II sub detector mock up . The mock up is composed by two main pieces, the beam pipe and the cooling block, arranged around the cooling block. A cooper ladder was placed in the outer layer, supported by the edges in two cooling blocks. In Figure 6 can be seen the distribution of these pieces. Both the cooling block and the Beam Pipe has internal refrigeration pipes. They are refrigerated with coolant cold down using a chiller. All this set-up is mounted in a big cage (figure 7) of methacrylate with ingress of dry air pipe and a controlled egress to allow the air going out, avoiding overpressure inside. The cage has some holes for the ingress of the Beam Pipe and cooling blocks cooling pipes being both supplies independent. Inside the cage there was a thermal-hygrometer to measure temperature and humidity

For the measurements several polyimide coated sensors where positioned inside the cage, all of them of Micronoptics [4].  One sensor was glued to the cooling pipe and one to the cooling block using adhesive tape. Another one was fixed using X60 to the cooling block. Three more were left free inside the cage to measure the temperature change and one was left outside to measure the room temperature.

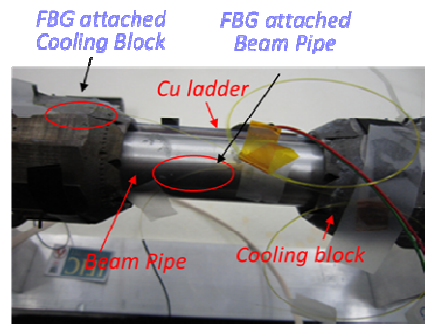

*Figure 6: PXD mock-up*

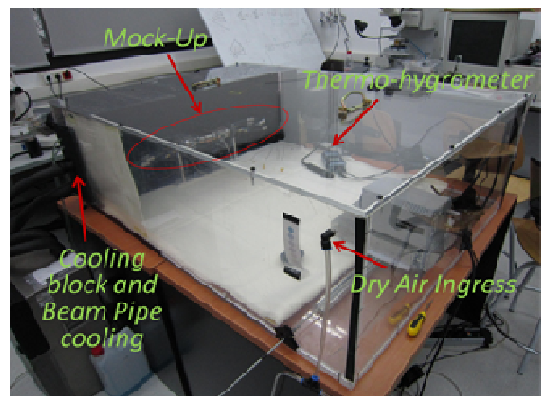

*Figure 7: Methacrylate cage set-up*

## 3.2   Belle II vertex detector Thermo-mechanical set up measurements

The set up was submitted to several cooling cycles to measure temperature and deformations. The cooling cycle of the set-up started with the reduction of the relative humidity inside the methacrylate cage. For this we dry air was injected inside the cage until the humidity inside the volume was below 8% . Then the chillers where switched on and the coolant was cooled up to a temperature of -20 º C. When the temperature in the PXD mock-up became stable, the chillers where switcher off  and  wait the time necessary for the mock temperature stabilization at room temperature.

In the figure 8 can be seen the wavelength change during two cooling cycles of the sensors glued by tape to the cooling block and the Beam Pipe. When the dry air supply was switched on, the sensor response changed with the humidity inside the cage. This effect is caused by the polyimide coating, which is hydrophilic [5] , [6]. The coating absorbs water swelling, causing an expansion in the fibre and the sensor. We have compared the FBG measurements with the thermo hygrometer´s ones (figure 9). As can be seen there is a good agreement between the tow measurements.

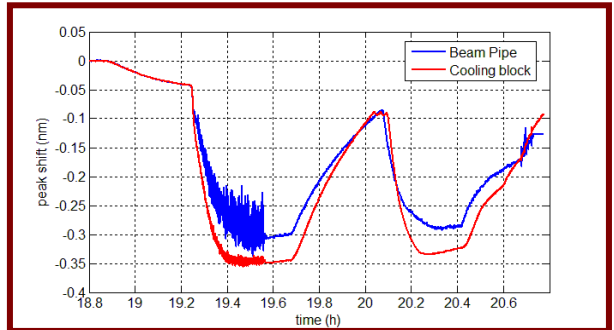

Figure 8. Peak shift during PXD mock-up two coolin cycles

The second step was to switch on the chillers of the beam Pipe and the Cooling block. It can be seen how the sensors wavelength change much faster than in humidity case. At soon as we start to cold down the set-up, appears a strange effect in the measurement done by FBG sensors, geater at lower temperatures. It is not a vibration, because it is not symmetric with respect to the nominal value, it goes always to higher values. This effect is caused by the dry air at room temperature being injected in the volume. This creates some temperature turbulences which FBG sensors are able to measure because its extremely low thermal inertia (only 250 μm of diameter). At the end of the first cooling we switched off the dry air and this effect disappeared. We measured another thermal cycle and we didn´t saw  this effect while dry air was switched off.

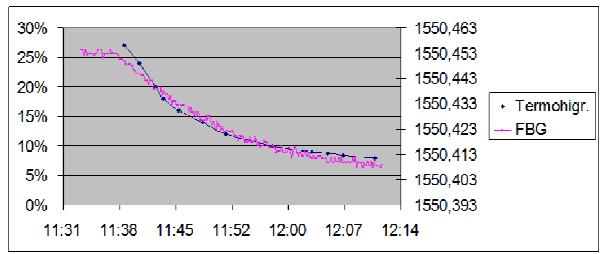

Figure 9. FBG – Thermohygrometer humidity measurements

With respect to the deformation measurements, in the figure 10 can be seen the wavelength change measured by the sensor fixed with X60 to the cooling block, and the one glued by kapton tape for the same cycle that in figure 9. The first thing to be addressed is that the X60 fixed sensor

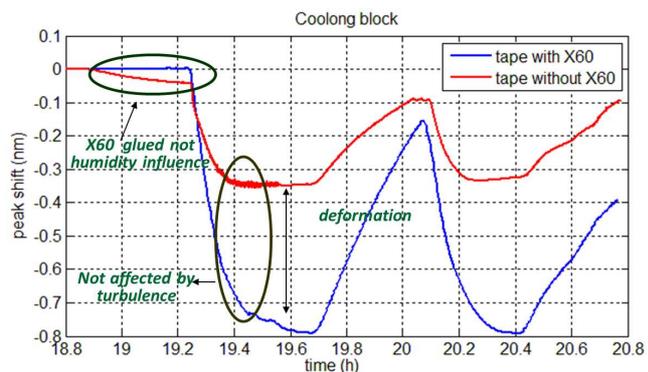

Figure 10. tape glued and X60 fixed sensors peak shift

*signal does not change with humidity and don´t see any strange effect during the cooling down of the mock-up[7]. This is because the sensor is protected by X60 from the atmosphere. The second thing to be addressed is that the wavelength change measured by the sensor fixed with X60 is much higher ( near the double) of the one measured by the sensor only glued with the adhesive tape. This is because this sensor measure temperature and X60 deformations at the same time.*

### 3.3 Belle II vertex detector Thermo-mechanical set up measurements conclusions

*We found that FBG sensors where very useful not only for environmental and deformation monitoring. They were also useful for diagnosing environmental conditions of the thermo-mechanical set up and to improve them.*

## 4 Conclusions

*Fiber optics sensors are the optimal technology to monitor environmental parameters, strain and vibrations and most important, Within the Aida FP7, FBG technology will be available to the interested community*